\documentclass[11pt,twoside,dvips]{article}
\usepackage{verbatim}
\usepackage{amssymb}
\usepackage{amsfonts}
\usepackage{amsmath}
\usepackage{pifont}
\usepackage[spanish,english]{babel}
\usepackage[pdftex]{graphicx}
\usepackage{verbatim}
\usepackage{epsfig}
\usepackage{bm}
\usepackage{longtable}
\usepackage{fancyhdr}
\usepackage[T1]{fontenc}
\usepackage[utf8]{inputenc}
\usepackage{authblk}

\begin{document}

\fancypagestyle{plain}{%
\fancyhf{}%
\fancyhead[LO, RE]{XXXVIII International Symposium on Physics in Collision, \\ Bogot\'a, Colombia, 11-15 september 2018}}

\fancyhead{}%
\fancyhead[LO, RE]{XXXVIII International Symposium on Physics in Collision, \\ Bogot\'a, Colombia, 11-15 september 2018}

\title{\vspace{-1cm} Solar Neutrino Measurements}
%\author{Author 1$\thanks{%
%e-mail: author1@inst}$, Author 2$\thanks{%
%e-mail: author2@inst}$, \\ Department, Institution, \\ Address}
%EndAName
\author[14]{A.~Pocar\thanks{pocar@umass.edu}}
\author[16]{M.~Agostini}
\author[16]{K.~Altenm\"{u}ller}
\author[16]{S.~Appel}
\author[7]{V.~Atroshchenko}
\author[24]{Z.~Bagdasarian}
\author[10]{D.~Basilico}
\author[10]{G.~Bellini}
\author[13]{J.~Benziger}
\author[9]{G.~Bonfini}
% ...

\affil[ ]{The Borexino Collaboration}

% ...
\affil[7]{National Research Centre Kurchatov Institute, 123182 Moscow, Russia}
% ...
\affil[9]{INFN Laboratori Nazionali del Gran Sasso, 67010 Assergi (AQ), Italy}
\affil[10]{Dipartimento di Fisica, Universit\`a degli Studi e INFN, 20133 Milano, Italy}
% ...
\affil[13]{Chemical Engineering Department, Princeton University, Princeton, NJ 08544, USA}
\affil[14]{Amherst Center for Fundamental Interactions and Physics Department, University of Massachusetts, Amherst, MA 01003, USA}
% ...
\affil[16]{Physik-Department and Excellence Cluster Universe, Technische Universit\"at  M\"unchen, 85748 Garching, Germany}
% ...
\affil[24]{Institut f\"ur Kernphysik, Forschungszentrum J\"ulich, 52425 J\"ulich, Germany}
% ...

\date{December 2, 2018}
\maketitle

\vspace{-0.8cm}
\begin{abstract}
We present the most recent results from the two currently running solar neutrino experiments, Borexino at the Gran Sasso laboratory in Italy and SuperK at Kamioka mine in Japan.
SuperK has released the most precise yet measurement of the $^8$B solar neutrino interaction rate, with a precision better than 2\%, consistent with a constant solar neutrino emission over more than a decade.
Borexino has released refined measurements of all neutrinos produced in the {\it pp} fusion chain.
For the first time, one single detector has measured the entire range of solar neutrinos at once. 
These new data weakly favor a high-metallicity Sun. 
Prospects for measuring CNO solar neutrinos with Borexino are discussed, and a brief outlook on the field provided.
\end{abstract}

\section{Solar neutrinos: the SuperK and Borexino experiments}
\begin{figure}[!t]
\begin{center}
\includegraphics[height=2.5in]{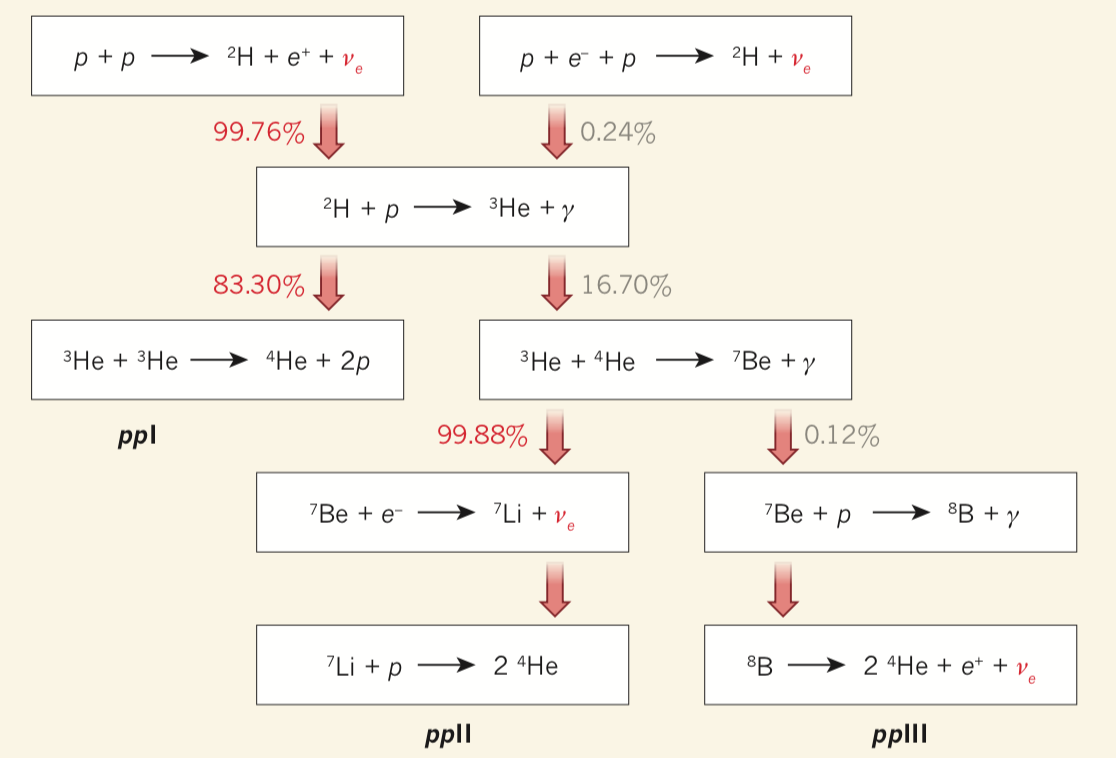}
\end{center}
\vspace{-0.4cm}\caption{The solar {\it pp} fusion chain.} 
\label{f:pp}
\end{figure}

\begin{figure}[!b]
\begin{center}
\includegraphics[height=1.6in]{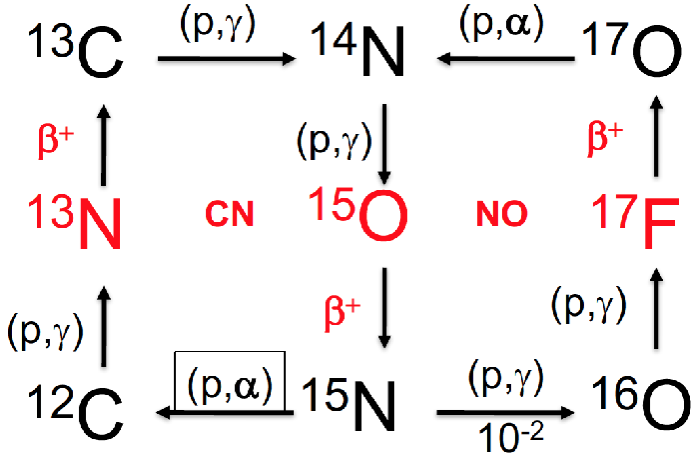}
\hspace{0.01in}
\includegraphics[height=1.7in]{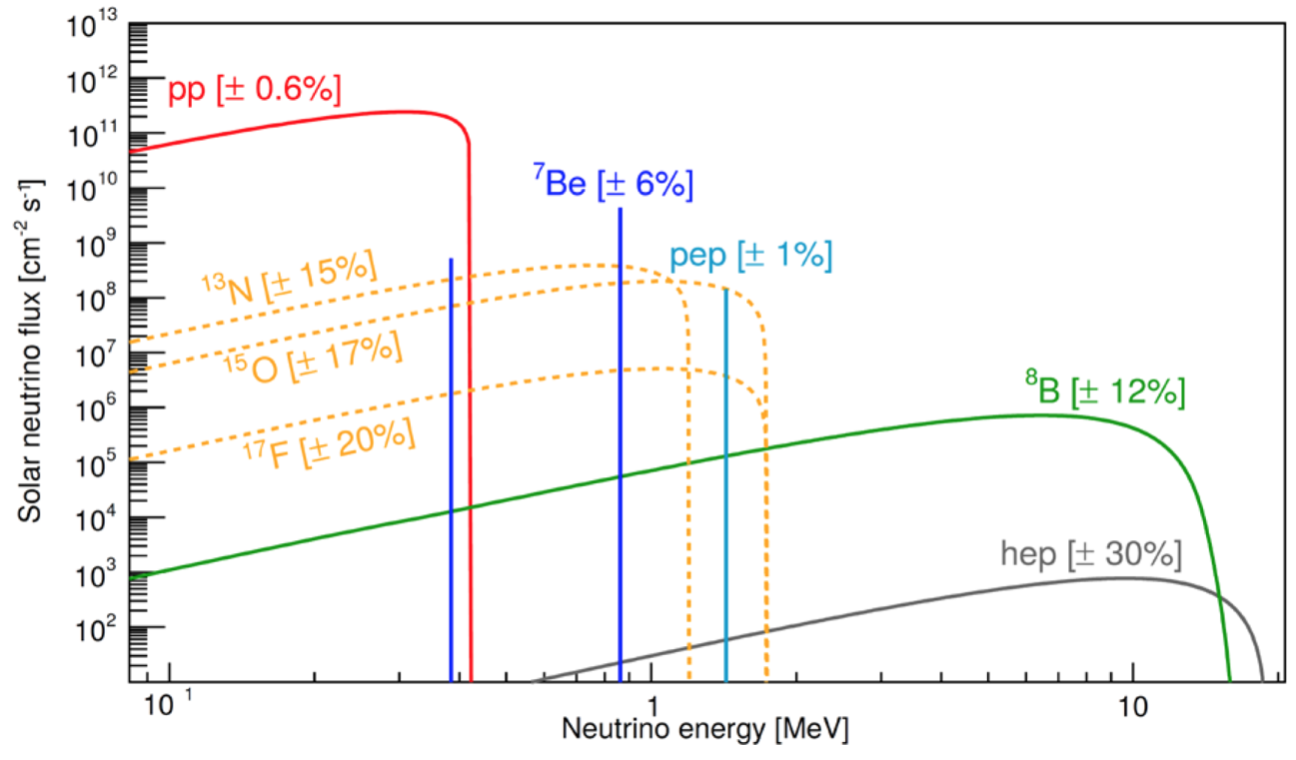}
\end{center}
\vspace{-0.4cm}\caption{Left: the CNO solar fusion cycle. Right: the SSM solar neutrino spectrum.} 
\label{f:cno-spectrum}
\end{figure}

The Sun is fueled by nuclear reactions fusing protons (hydrogen) into helium via a set of reactions summarized as:
$$
4p\,\longrightarrow\,^4He+2e^+ +2\nu_e + (24.7 + 2m_ec^2)[MeV].
$$
In the Sun, 99\% of the times this process is carried out through a set of reactions known as the {\it pp}-chain, initiated by the fusion of two protons as illustrated in Fig.~\ref{f:pp}. 
Fig.~\ref{f:cno-spectrum} shows the reactions believed to contribute the remaining $\sim$1\%, in which proton fusion is catalyzed by heavier elements, enhanced by higher {\it metallicity} (in astrophysics, all elements heavier than helium are call {\it metals}). 
Also shown is the spectrum of solar neutrinos predicted by the Standard Solar Model (SSM).
A comprehensive review of solar neutrino physics, with connections to their experimental investigation, their role in the discovery of neutrino oscillations, and the definition of neutrino flavor conversion parameters is found in~\cite{Haxton:2013cx}.

SuperK is a massive detector using 50 ktons of ultra-pure water as target for solar neutrinos.
Only $^8$B neutrinos can transfer enough energy to the electrons they scatter off to produce \v{C}erenkov light, read out by $\sim$10,000 photomultiplier tubes.
The \v{C}erenkov technique retains some directional information of incoming neutrinos, at a price of a relatively high energy threshold.
SuperK has just concluded its period SK-IV period of data taking and has been running since 1996.

Borexino measures solar (and other low energy) neutrinos interacting with a spherical target of $\sim$300 tonnes of organic liquid scintillator, separated by surrounding buffer fluid by a thin, transparent nylon vessel.
Scintillation pulses from neutrino interactions, as well as other (mostly background) ionizing events are detected by $\sim$2,000 8-inch photomultiplier tubes (PMTs).
Borexino displays the lowest energy threshold for solar neutrino detection to date.
It, however, lacks to ability to retain directional information of neutrino interactions.
Borexino has been running since 2007 and is completing its Phase-2 data taking period.

Here we present an update on the results from SuperK Phase IV (Sec.~\ref{s:superk}, last reported on in 2016~\cite{Abe:2016el}.
We also report on recent measurements from Borexino Phase-2 that cover the entire {\it pp} chain (Sec.~\ref{s:bx}).
These results improve on earlier measurements, mostly made with Phase-1 data~\cite{Bellini:2014un,Collaboration:2014th}. 
We provide a status update on the prospects of measuring CNO solar neutrinos with Borexino (Sec.~\ref{s:cno}), and offer a brief outlook on the solar neutrino sub-field (Sec.~\ref{s:outlook}).

\begin{figure}[!t]
\begin{center}
\includegraphics[height=2in]{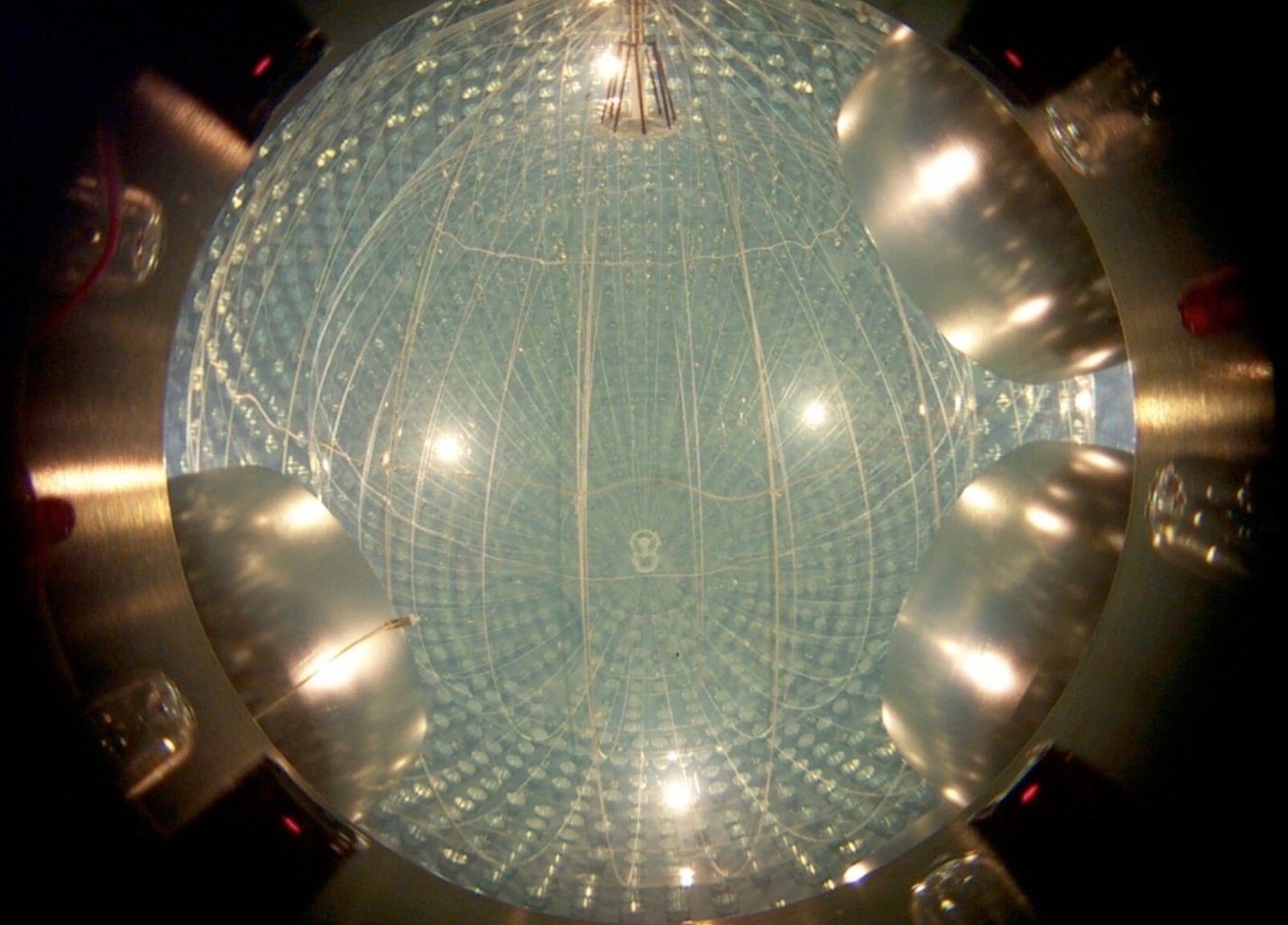}
\hspace{0.1in}
\includegraphics[height=2.1in]{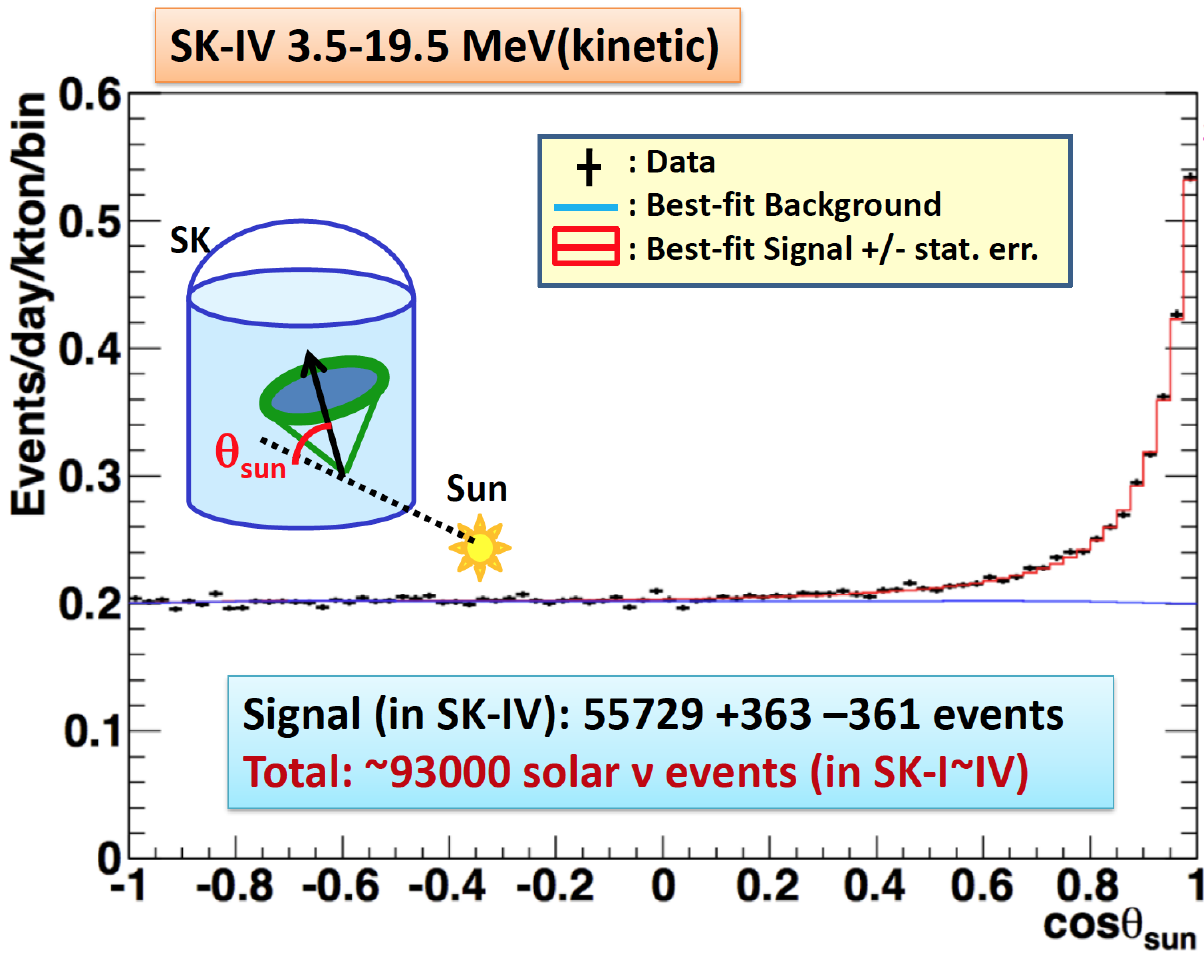}
\end{center}
\vspace{-0.4cm}\caption{Left: The Borexino inner detector filled with purified scintillator in 2007. The Borexino scintillator is arguably the most radio-pure bulk material ever measured. Right: Reconstructed direction of incidence of $\sim$100,000 solar neutrinos detected by SuperK periods I-IV, clearly showing the Sun's direction (the plot is courtesy of Yasuo Takeuchi, who presented it at RICH18).} 
\label{f:bx-superk}
\end{figure}

\section{Results from SuperK SK-IV}
\label{s:superk}

\begin{figure}[!t]
\begin{center}
\includegraphics[height=2.1in]{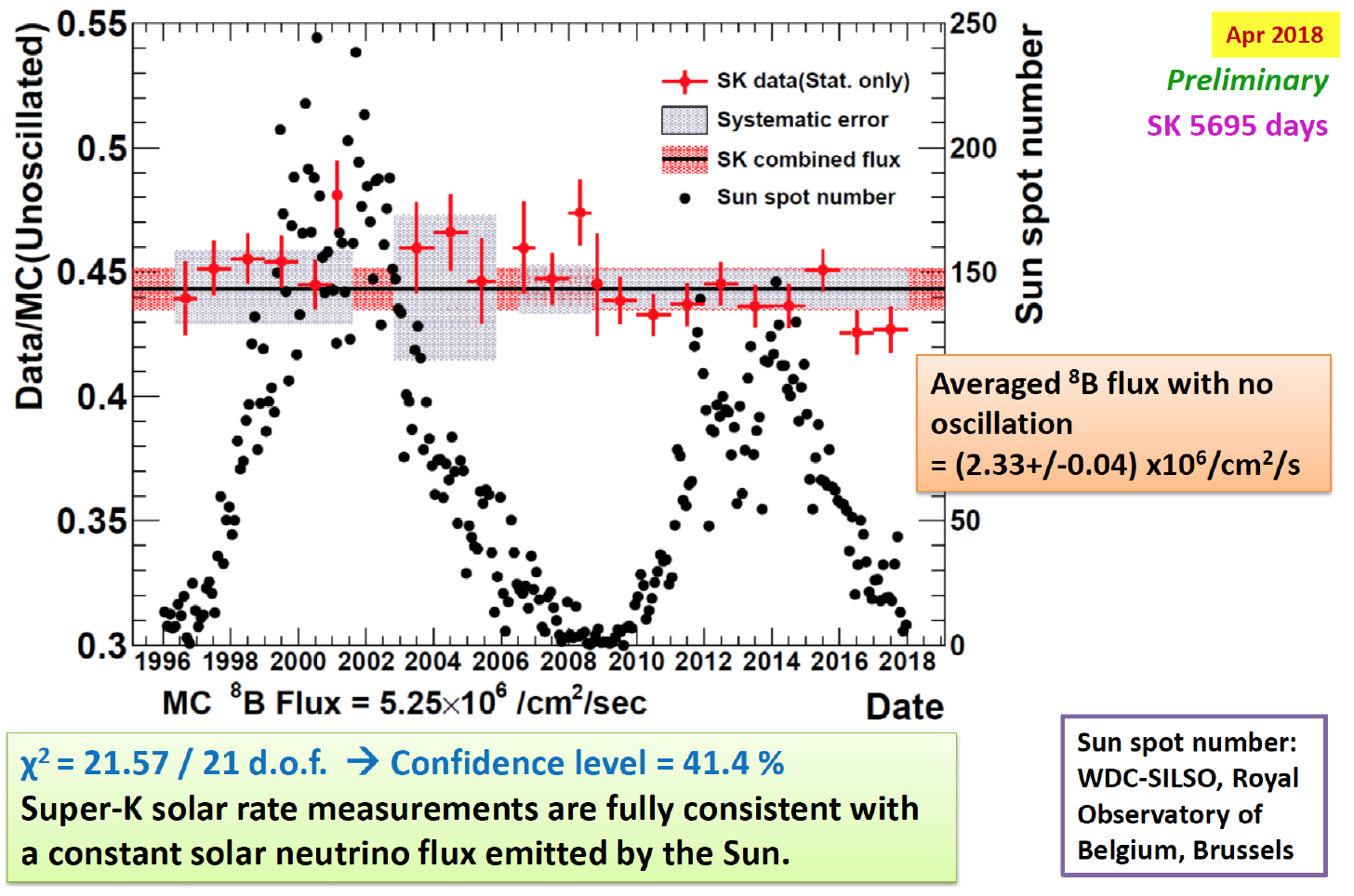}
\hspace{0.1in}
\includegraphics[height=2.1in]{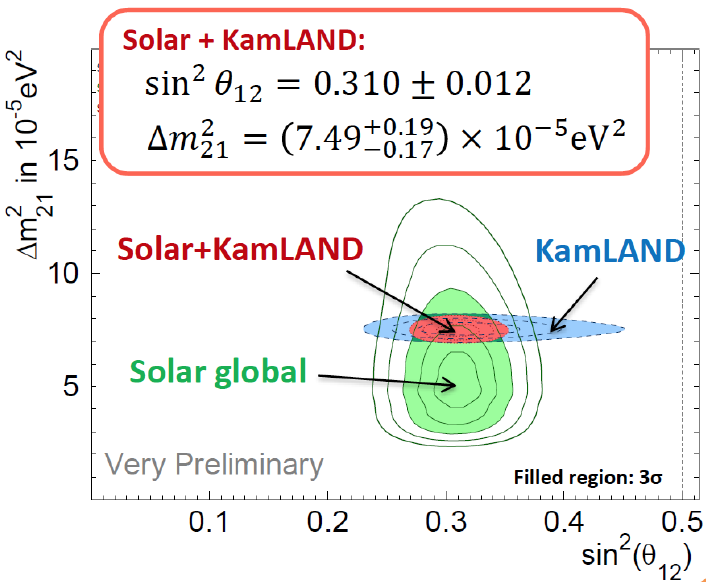}
\end{center}
\vspace{-0.4cm}\caption{Left:  Right: Precision measurement of the $^8$B solar neutrino interaction rate in SuperK, and absence of its seasonal variation on- (or off-) phase with solar spot activity (plots are courtesy of Yasuo Takeuchi at RICH18). Right: very preliminary 'solar only' measurement of the $\Delta^2_{12}$ neutrino oscillation parameter including all SuperK data.} 
\label{f:superk}
\end{figure}

SuperK has run in its SK-IV period since 2008, for 2,860 days of live time.
Improvements from the preceding SK-III period focused on an upgrade of the electronics, which allowed the energy threshold for solar neutrino physics to be reduced from 4.5 MeV to 3.5 MeV of electron recoil energy, taking full advantage of the large optical coverage offered by the 11,129 PMTs.
The ability of SuperK to determine the solar origin by directional information is illustrated in Fig.~\ref{f:bx-superk}.
The detection of solar neutrinos with strong statistical power is enabled by the massive size of the SuperK detector.

Due to its relatively high energy threshold (although it is a challengingly low threshold for the \v{C}erenkov technique), SuperK is only sensitive to $^8$B (and, if there, $hep$) solar neutrinos.
It, however, able to measure them with exquisite precision.
The newly-released averaged $^8$B solar neutrino flux from SK-IV assuming no neutrino oscillations is $(2.33\pm 0.04)\times 10^{-6}/{\rm cm^2/s}$, a precision of 1.7\%.
In addition, no time-varying interaction rate of these neutrinos on- (or off-) phase with solar spot activity is observed (see Fig.~\ref{f:superk}, right panel).
Using the entire data set (periods I-IV) SuperK presents a very preliminary SuperK-only measurement of the solar neutrino mass splitting $\Delta^2_{12}$, finding a value which is 2$\sigma$ from that measured with higher precision by KamLAND (see Fig.~\ref{f:superk}, right panel). 

The next phase for SuperK is SK-Gd.
For this phase, the water target is being doped with gadolinium to greatly enhance neutron detection, with the aim to measure diffused supernov{\ae} neutrinos (SK-Gd would also have enhanced sensitivity to neutrinos from individual galactic supernov{\ae}, should one explode in the near future).

\section{Results from Borexino Phase-2}
\label{s:bx}
\begin{figure}[!t]
\begin{center}
\includegraphics[height=1.8in]{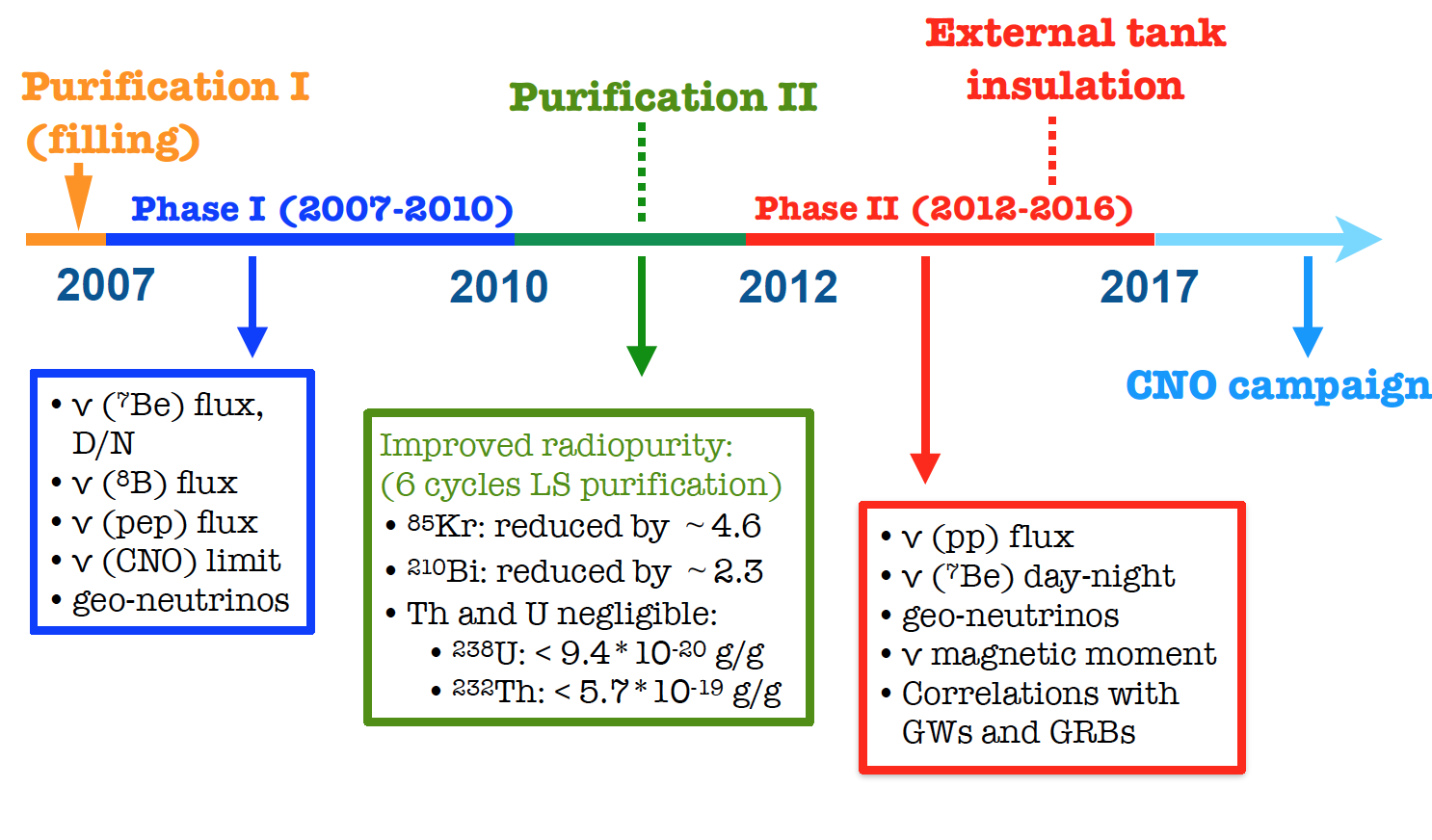}
\hspace{0.1in}
\includegraphics[height=1.8in]{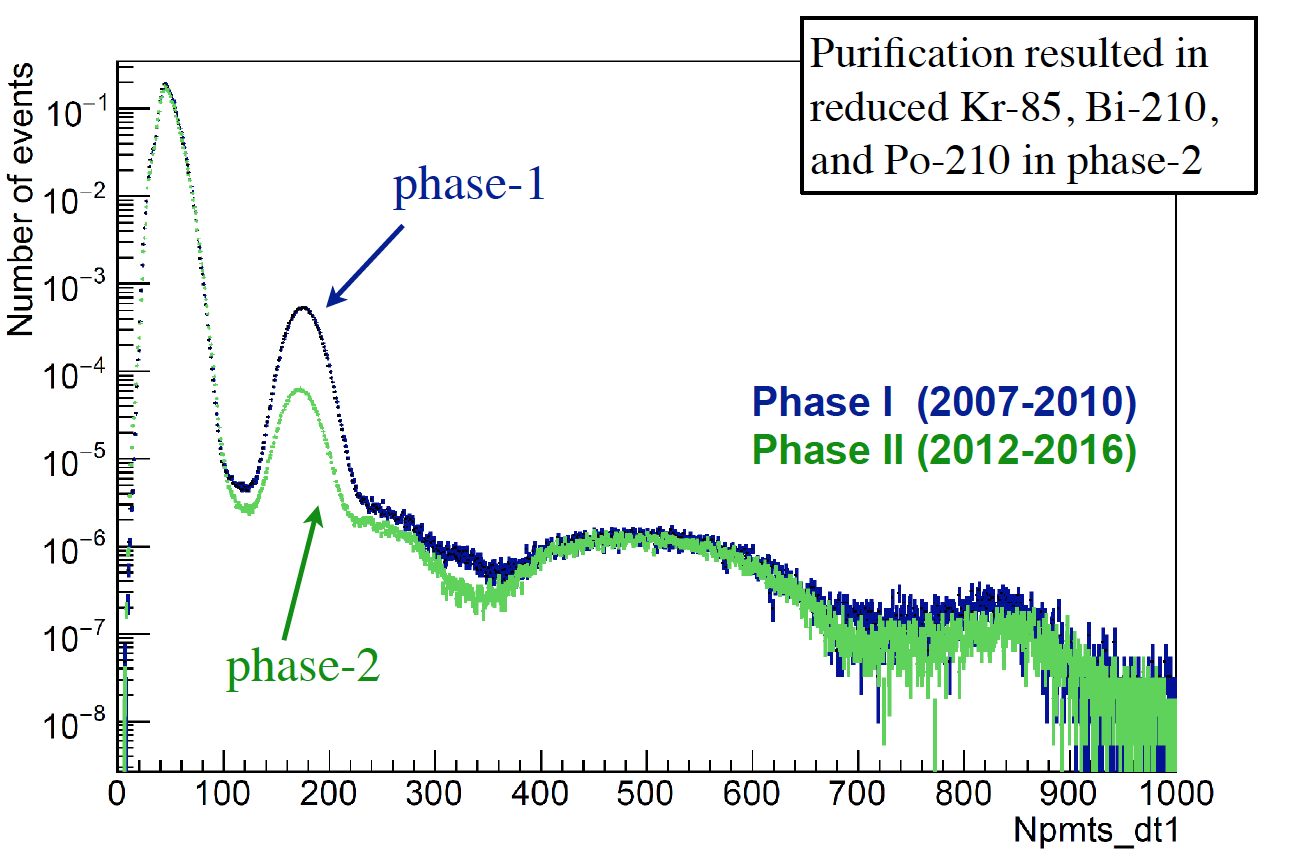}
\end{center}
\vspace{-0.4cm}\caption{Left: A summary of the Borexino timeline. Right: Comparison between Borexino Phase-1 and Phase-2 data, highlighting $^{85}$Kr and $^{210}$Bi background reduction.} 
\label{f:bx1}
\end{figure}

Borexino has been running its Phase-2 science run since 2012 (see Fig.~\ref{f:bx1}, left panel).
This phase followed an extended purification period in which the scintillator underwent six cycles of water extraction with partial-vacuum nitrogen stripping.
Purification achieved a $\sim$ 5-fold reduction of $^{85}$Kr and a $\sim$ two-fold reduction of $^{210}$Bi (see Fig.~\ref{f:bx1}, right panel).
In addition, it further reduced $^{232}$Th and $^{238}$U equivalent background to unprecedented low levels, with limits set for both isotopes of $<5.7\times10^{-19}$ g/g and $<9.4\times10^{-20}$ g/g, respectively.

\begin{figure}[!b]
\begin{center}
\includegraphics[height=2.7in]{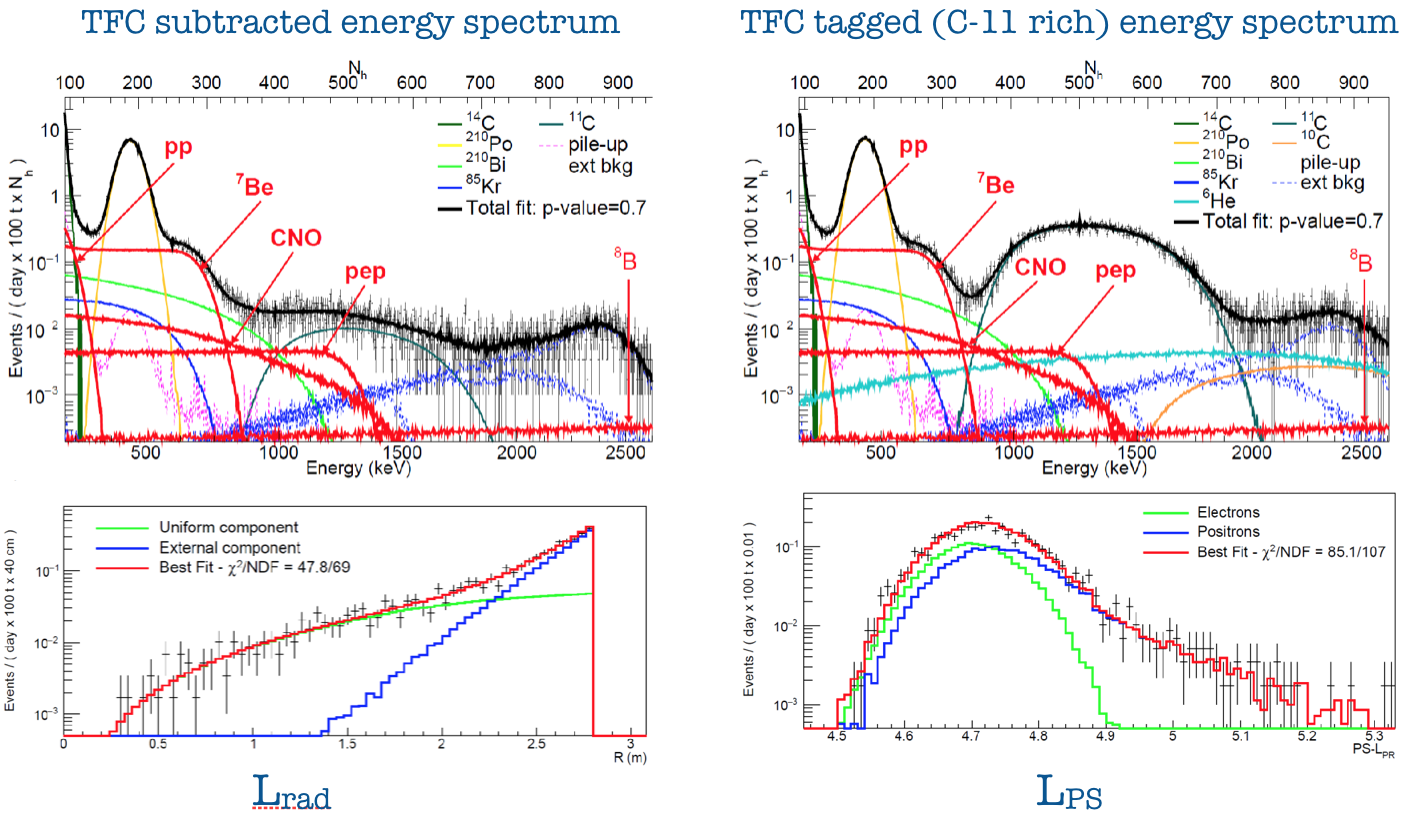}
\end{center}
\vspace{-0.4cm}\caption{Illustration of the Phase II Borexino fitting strategy used to simultaneously determine the interaction rate of {\it pp}, $^7$Be, and {\it pep} solar neutrinos. The fit also returns an upper limit for CNO neutrinos, as well as rates for all relevant background sources.} 
\label{f:bx-c11}
\end{figure}

Phase-2 Borexino data provided the opportunity to improve the precision measurements of all solar neutrino fluxes~\cite{BorexinoCollaboration:2018bt}. 
The improved radio-purity of the scintillator was one key ingredient for a measurement of $^7$Be neutrinos with a better than 3\% precision.
This is twice as small than the theoretical uncertainty from the SSM, and can be used, in combination with assumptions on the solar neutrino oscillation parameters, to better our understanding of the sun.
Another ingredient is a greatly-improved Monte Carlo simulation package was developed~\cite{Agostini:2018bu} that allowed a more accurate determination of the energy response over a wide energy range, and background-suppressing analysis tools were refined, adding to the intrinsic improvement provided by an extended data set.
A multi-variate approach was used to identify and suppress the cosmogenic $^{11}$C background ($\sim$30 minute half life)), a $\beta^+$ emitter covering the energy range relevant for {\it pep} and CNO neutrino detection.
Positron emission with the production (50\% of the times) of 3 ns-lived ortho-positronium and the production of annihilation gamma rays (extended and with a slightly different ionization density profile in the scintillator) produces a statistically distinguishable time profile of the scintillation pulse from that of electron events.
A likelihood was built using such a pulse shape parameter, the radial distribution of events, and the simultaneous fit of $^{11}$C-rich and $^{11}$C-subtracted energy spectra.
This procedure for determining neutrino fluxes is illustrated in Fig.\ref{f:bx-c11}, and allowed us to simultaneously fit the Borexino data between $\sim$200 keV and $\sim$2.5 MeV, including the interaction rate of {\it pp}, $^7$Be, {\it pep}, and CNO solar neutrinos and the overlapping backgrounds (previous measurements were carried out focusing on narrower energy regions).

\begin{figure}[!t]
\begin{center}
\includegraphics[height=1.5in]{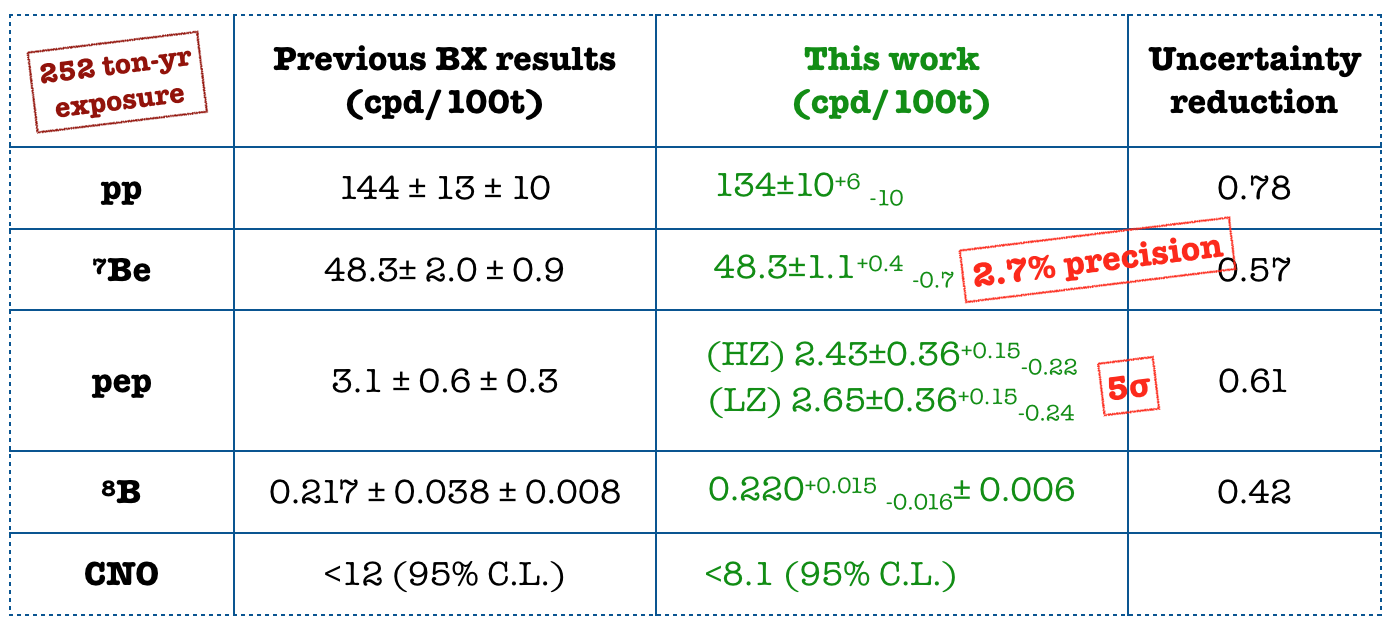}
\hspace{0.1in}
\includegraphics[height=1.5in]{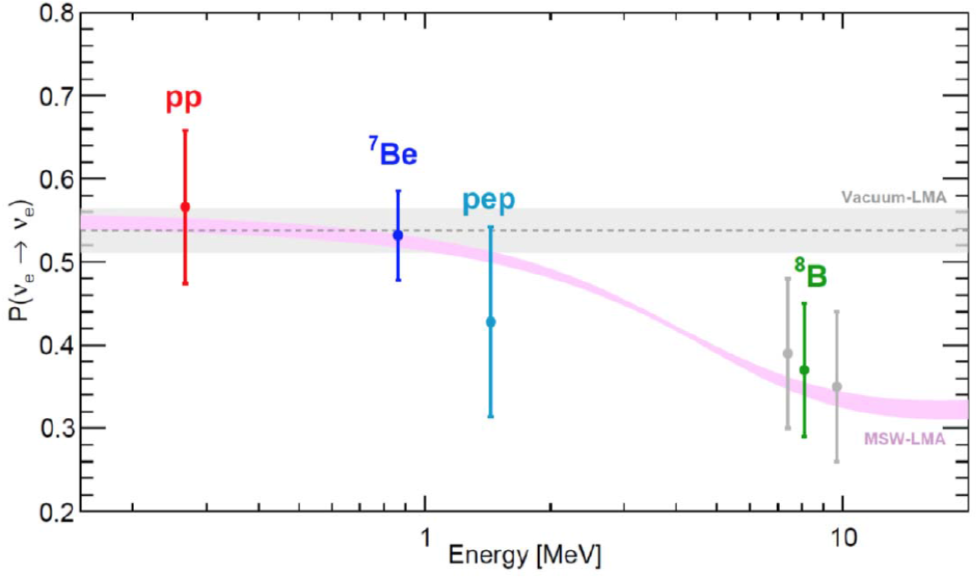}
\end{center}
\vspace{-0.4cm}\caption{Left: Phase-II Borexino solar neutrino results compared to pre-2018 measurements. Right: The electron flavor survival probability of the solar neutrino components measured with Borexino. The pink band represents the energy-dependent survival probability under the assumption of HZ SSM.} 
\label{f:bx2}
\end{figure}

The Phase II results are collected in the table in Fig.~\ref{f:bx2}. 
The uncertainty on the {\it pp} neutrino rate is reduced to $\sim$10\% compared to the first measurement~\cite{Collaboration:2014th}.
This is no small feat, given that its very measurement went beyond what Borexino had proposed to do in the beginning and that several experiments have been proposed over time whose only or main goal was the study of {\it pp} neutrinos.
The {\it pep} neutrinos are definitively discovered with 5$\sigma$ significance, and a tight limit of $<8.1$ counts per day (cpd)/100t (95\% C.L.) is set for the CNO neutrino rate.

The $^8$B solar neutrino rate was measured separately using the same data set.  
The entire scintillator volume was used (\emph{i.e.} no fiducial cut), a choice dictated by the need to boost the statistics for this dimmer solar neutrino component.
A lower energy cut of 3.2 MeV was placed, above much of the natural radioactivity but retaining a large fraction of the $^8$B solar neutrino signal.
The $^8$B rate is not as precise as that measured by SuperK, but it is compatible with it and it is measured with the lowest energy threshold of all experiments for this component.

The Borexino Phase II neutrino fluxes allow us to attempt at addressing the open issue of solar metallicity, {\it i.e} the abundance of elements heavier than helium. 
High- and low-metallicity solar models (referred to as HZ and LZ, respectively) result from contrasting measurements. 
Helio-seismological data prefer HZ solar photospheric abundances. 
Solar metallicity affects solar neutrino fluxes, most prominently that of CNO neutrinos with a $\sim$30\% higher flux predicted by the HZ model compared to the LZ one.
However, both the $^7$Be and $^8$B fluxes are $\sim$10\% higher in the HZ SSM, while the {\it pp} and {\it pep} fluxes are higher for the LZ SSM~\cite{Haxton:2013cx}.
Thus the solar metallicity question could be addressed with high-precision measurements of these fluxes.

The $^7$Be and {\it pp} neutrino fluxes can be used to compare the relative weight of the two helium-helium fusion reactions by computing 
$$
R \equiv \frac{^3 {\rm He}+ ^4{\rm He}}{^3{\rm He} + ^3{\rm He}} = \frac{2\phi(^7{\rm Be})}{\phi(pp)+\phi(^7{\rm Be})},
$$
which is predicted to be ($0.180\pm0.11$) and  ($0.161\pm0.10$) for HZ and LZ SSM, respectively.
This ratio measured by Borexino is $R_{BX}=0.178^{+0.027}_{-0.023}$.

\begin{figure}[!h]
\begin{center}
\includegraphics[height=1.8in]{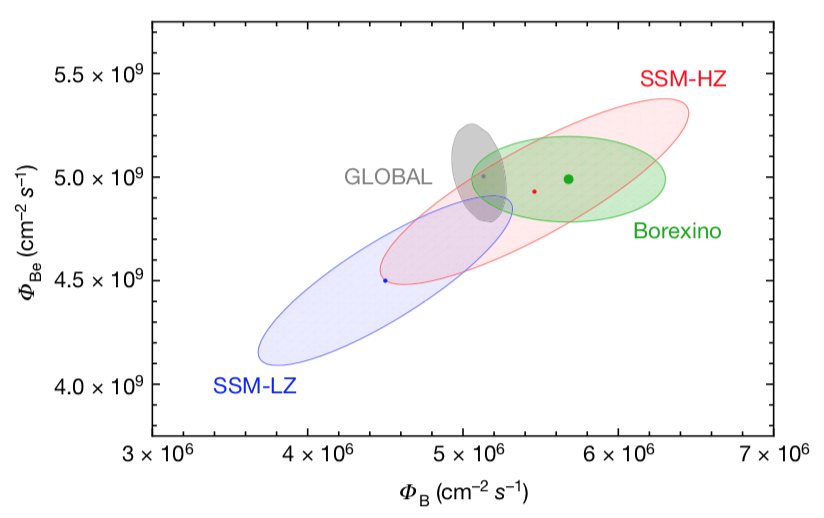}
\hspace{0.01in}
\includegraphics[height=1.8in]{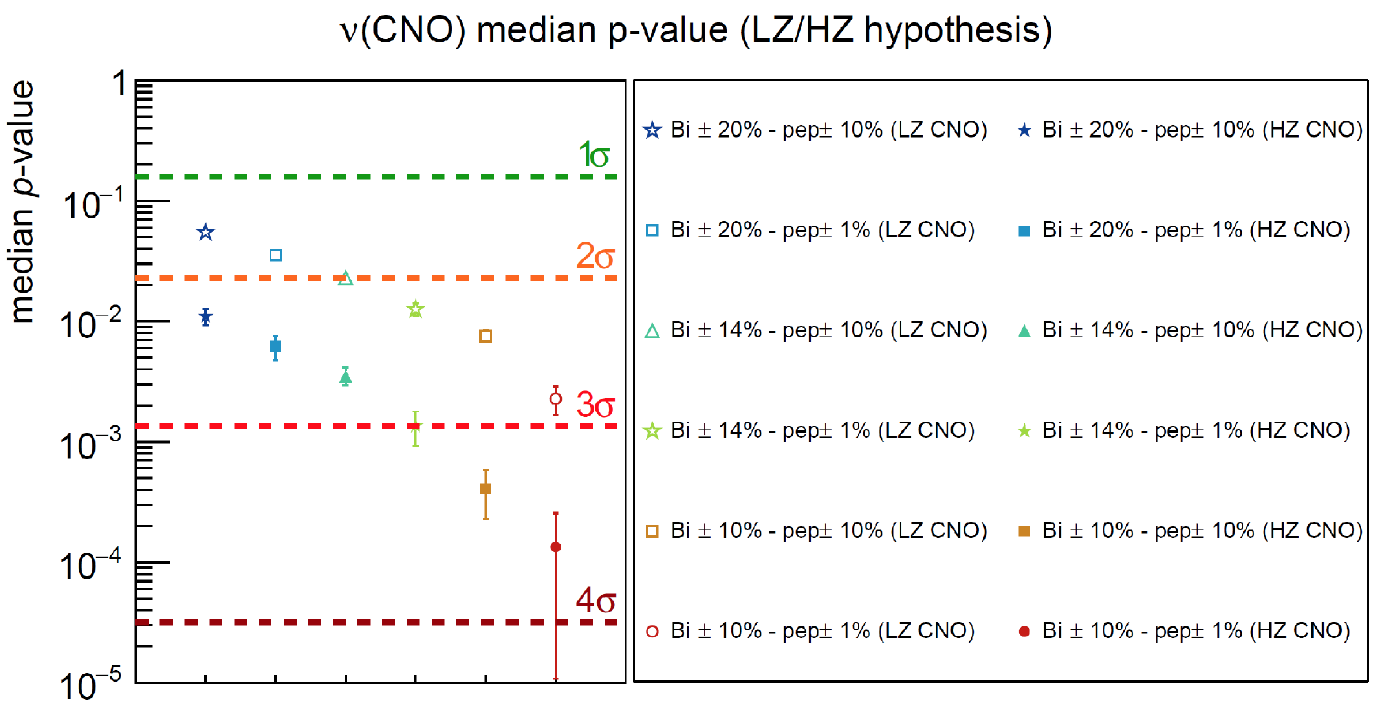}
\end{center}
\vspace{-0.4cm}\caption{Left: Phase-II Borexino and solar neutrino experiments $^7$Be and $^8$B solar neutrino results compared against theoretical predictions from HZ and LZ SSM. Ellipses indicate 1 $\sigma$ contours. Right: Borexino discovery potential for CNO solar neutrinos. The study shows that $^{210}$Bi background needs not only be low, but known at the 10-20\% level.Borexino discovery potential for CNO solar neutrinos. The study shows that $^{210}$Bi background needs not only be low, but known at the 10-20\% level}
\label{f:ellipses}
\end{figure}

Similarly, the measured $^7$Be and $^8$B can be used together and compared with SSM HZ and LZ cases, as shown in Fig.~\ref{f:ellipses}. 
Borexino data alone mildly prefer the HZ SSM. 
This hint is weakened by including all solar neutrino data, which notably provides a more precise value for the $^8$B neutrinos as measured by the SuperK experiment, in the analysis.
In addition, theoretical uncertainties barely differentiate the two scenarios in this case.

\section{Measuring CNO neutrinos}
\label{s:cno}
Of great astrophysical interest is the measurement of CNO neutrinos, as they are arguably the most direct probe of solar metallicity.
In order to assess how far Borexino is from measuring CNO neutrinos, the current Borexino limit $<8.1$ cpd/100t (95\% C.L.) should be compared with SSM HZ and LZ predictions of  $4.91\pm0.52$ cpd/100t and $3.52\pm0.37$ cpd/100t, respectively.
On one hand, this extra factor of two seems at arm's reach of the experiment. 
On the other, one background exists that poses a serious challenge, the $\beta$ emitter $^{210}$Bi.

\begin{figure}[!h]
\begin{center}
\includegraphics[height=1.7in]{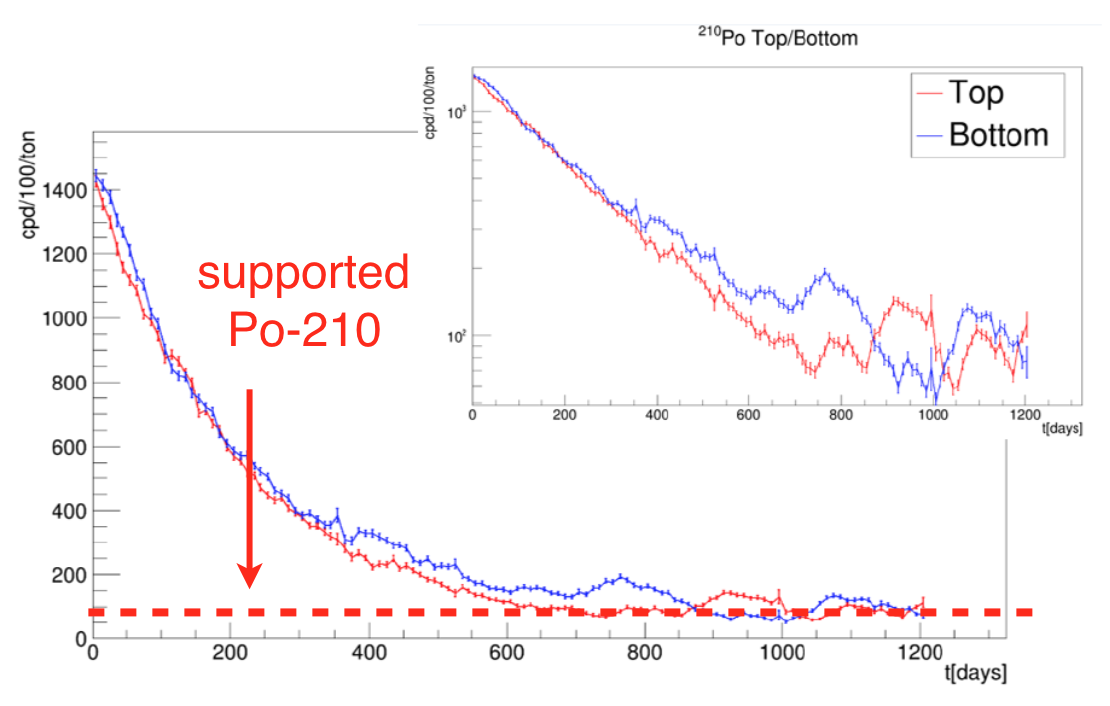}
\hspace{0.01in}
\includegraphics[height=1.7in]{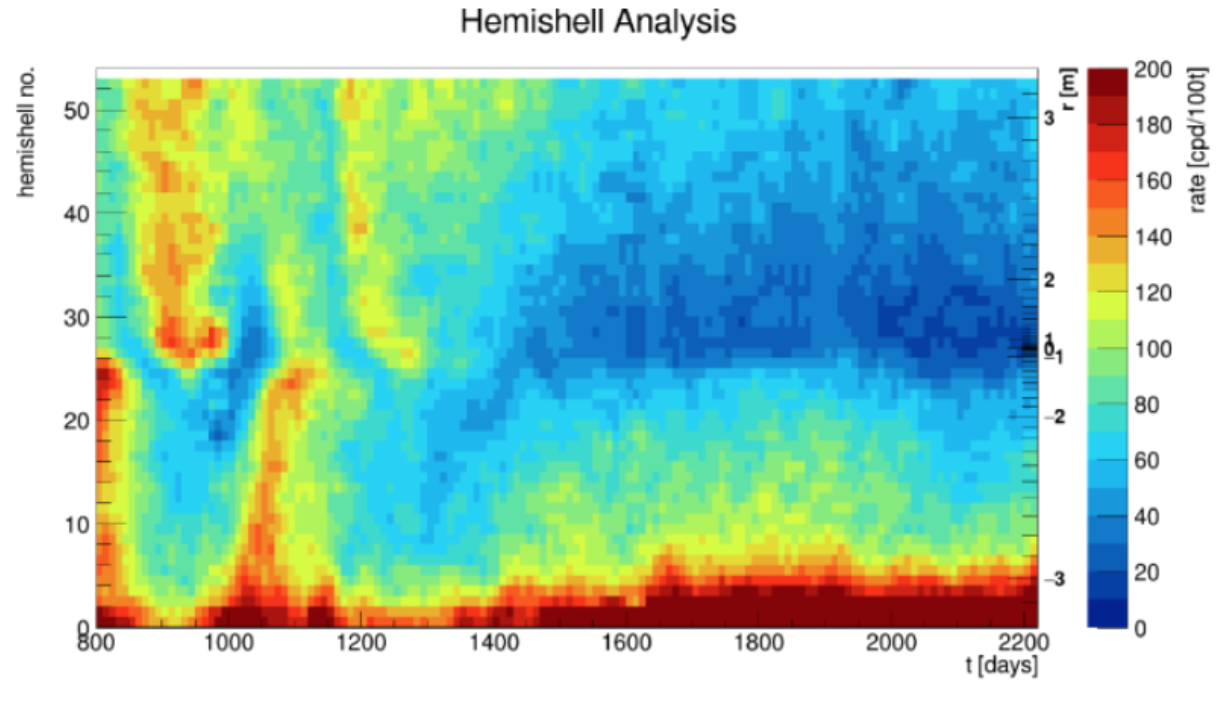}
\end{center}
\vspace{-0.4cm}\caption{Left: The $^{210}$Po activity in the inner part of the Borexino scintillator volume versus time for the Phase II dataset. 
The data is divided between top and bottom half of the volume. It shows deviations from a pure exponential decay with the $^{210}$Po half life (138 days), a behavior consistent with 'dirtier' scintillator from outside the fiducial volume entering and leaving it as a result of convective motions. 
Right: the effectiveness of thermally insulating the Borexino tank is evident from the ensuing layering of the $^{210}$Po activity inside the fiducial volume.} 
\label{f:po210}
\end{figure}

$^{210}$Bi is a decay product of $^{222}$Rn, and because of this is found in air and on virtually all surfaces. 
It is sustained by its long-lived $^{210}$Pb, and is followed by a relatively long-lived $\alpha$ emitter, $^{210}$Po.
The $^{210}$Bi $\beta$ spectrum is quasi-degenerate with that of electrons recoiling off CNO neutrinos, as shown in Fig.~\ref{f:bx-c11}.
Detection of CNO neutrinos thus hinges on having very low and well measured $^{210}$Bi background. 
One could determine the $^{210}$Bi activity by measuring the supported $^{210}$Po component after the fraction which is out of equilibrium has decayed away, as proposed in Ref.~\cite{Villante:2011wj}.
Fig.~\ref{f:po210} (left panel) shows the $^{210}$Po activity in the Borexino fiducial volume versus time for approximately the past three years.
A precise determination of the steady-state component is made difficult by background fluctuations caused by scintillator mixing due to convective motions with timescales of several months.
In 2015, the entire Borexino detector was thermally insulated from the air of the experimental hall, the effect of which can be appreciated in Fig.~\ref{f:po210} (right panel).
The collaboration is looking hard into whether, with this stabler detector, $^{210}$Bi is low and constrained enough to allow for a measurement of CNO neutrinos in the near future.
Fig.~\ref{f:ellipses} (right panel) shows a detection sensitivity study for CNO solar neutrinos, which shows that $^{210}$Bi background needs to be be both low and precisely measured for a detection claim to be possible.

\section{Outlook}
\label{s:outlook}
SuperK has been running for more than two decades, measuring solar $^8$B neutrinos with great precision and contributing to the definition of neutrino oscillations in the solar sector.
Borexino has been running for more than ten years, measuring the entire solar neutrino energy spectrum with increasing precision. Noteworthy is the measurement of $^7$Be neutrinos with better that 2.7\% precision, a prominent piece of the "solar neutrino puzzle" and directly contributing to the refinement of solar models. 

In the next few years, Borexino will attempt to measure the CNO solar neutrino interaction rate, which could resolve the solar metallicity puzzle, an important open question about the sun.

Looking ahead, solar neutrino physics does not have many projects lined up to continue it.
The SNO+ experiment in Canada could measure CNO neutrinos better than Borexino if they are able to achieve similar levels of radio-purity, since it is 3 times larger and much deeper (with much lower $^{11}$C background).
However, their priority is a program to measure neutrinoless double beta decay of $^{130}$Te, which looks incompatible with a concurrent solar neutrino program.
The JUNO experiment in China would display a much larger target than Borexino, but it is shallower and has to prove scintillator radio-purity can match that of Borexino~\cite{Salamanna:2018vp}.
A letter of intent has been submitted for a liquid scintillator experiment at Jinping laboratory in China.
This would be a deep, very large (2 kton fiducial mass) experiment with Borexino-like radio-purity, which could measure most solar neutrino components with percent precision~\cite{Beacom:2016vl}.
A 300 tonne liquid argon detector, such as that imagined by the DarkSide collaboration, could perform high precision spectroscopy of $pep$, $^7$Be, and CNO neutrinos~\cite{Franco:2016ex}.
The much larger, yet higher energy threshold liquid argon DUNE experiment has shown competitive sensitivity for  $^8$B 
solar neutrinos, including day/night modulations, and could be sensitive enough to observe $hep$ neutrinos~\cite{Capozzi:2018tz}.

\section*{Acknowledgments}
The Borexino Collaboration acknowledges the generous hospitality and support of the Laboratori Nazionali del Gran Sasso (Italy).
The Borexino program is made possible by funding from INFN (Italy), NSF (USA), BMBF, DFG (OB168/2-1, WU742/4-1, ZU123/18-1), HGF, and MPG (Germany), RFBR (Grants 16-02-01026 A, 15-02-02117 A, 16-29-13014 ofim, 17-02-00305 A), RSF (Grant 17-02-01009) (Russia), and NCN (Grant UMO 2013/10/E/ST2/00180) (Poland).

%\appendix
%\section{app 1}

\bibliographystyle{SciPost_bibstyle}
\bibliography{pic2018}
%\begin{thebibliography}{99}
%\bibitem{t} A reference
%\end{thebibliography}
%

\end{document}